\documentclass[11pt]{article}

\usepackage{amsmath}
\usepackage{float}
\usepackage{upgreek}
\usepackage{booktabs}
\usepackage{graphicx}

\begin{document}

% Full title of the paper (Capitalized)
\title{Topology of Platonic Spherical Manifolds: From Homotopy to Harmonic Analysis}

% Authors (Add full first names)
\author{Peter Kramer\\
Institut f\"ur Theoretische Physik\\
der Universit\"at T\"ubingen\\ 
T\"ubingen 72076, Germany}

% Abstract (Do not use inserted blank lines, i.e. \\) 

\maketitle

\begin{abstract}
We carry out the harmonic analysis on four Platonic spherical three-manifolds with different topologies.
Starting out from the homotopies (\textit{Everitt 2004} [4]), we convert them into deck operations, 
acting on the simply connected three-sphere as the cover, and obtain the corresponding variety of deck groups. 
For each topology, the three-sphere is tiled into copies of a~fundamental domain under the corresponding deck group. 
We employ the point symmetry of each Platonic manifold to construct its fundamental domain as a~spherical orbifold. 
While the three-sphere supports an~orthonormal complete basis for harmonic analysis formed by Wigner polynomials,
a given spherical orbifold leads to a~selection of a~specific subbasis.
The resulting selection rules find applications in cosmic topology, probed by the cosmic microwave background.
\end{abstract}

\section{Introduction}

In 1917, Einstein replaced the Euclidean three-space by the three-sphere and so introduced the first spherical manifold to describe the spatial part of the Universe~\cite{EI17}.
More generally, three-manifolds with non-Euclidean topology have in recent years found applications in cosmology, 
in particular in relation to the multipole analysis of the cosmic microwave background (CMB) radiation.
Cosmological models with positive curvature are related to the three-sphere.

The three-sphere covers all spherical three-manifolds, including the Platonic manifolds considered here.
For the Poincar\'{e}s dodecahedral three-manifold, the first orthonormal basis required for the multipole analysis was constructed in~\cite{KR05}, using Lie algebraic methods and drawing on work by Klein 1884~\cite{KL93}.

Compared to other manifolds, the Platonic three-manifolds have the advantage that their homotopies have been derived and classified in 2003 by Everitt~\cite{EV04}.
The homotopies are fundamental for the mathematical analysis of the topology of manifolds~\cite{SE34,WO84,TH97,RA94,MO87}.
The connection of multipole-resolved CMB measurements to cosmic topology is provided by selection rules for the harmonic analysis.

Surprisingly, there is no systematic account that links the homotopy of Platonic manifolds to deck groups and to their harmonic analysis and selection rules. 
Our aim is to close this gap.

We start by relating the three-sphere to the group $SU(2,C)$ in Section~\ref{sec:3sphere}.
The representations of the group $SU(2,C)$ yield the Wigner polynomials introduced in Section~\ref{sec:wigner}, with unitary actions resulting in a~multipole decomposition discussed in Section~\ref{sec:unitary}.
In Section~\ref{sec:grids}, we index the Wigner basis polynomials by points on an~$m$-grid in two dimensions
in preparation for deriving topological selection rules.

The Platonic spherical polyhedra are constructed in Sections~\ref{sec:coxeter} and~\ref{sec:unimodular} 
by use of four spherical Coxeter groups for the tetrahedron, cube, octahedron and dodecahedron.
The Platonic homotopy groups from~\cite{EV04} are expressed in Sections~\ref{sec:homotopy} and~\ref{sec:synopsis} by the gluings of faces and directed edges of the Platonic polyhedra.

In the central Section~\ref{sec:deck}, we use the isomorphism of homotopy and deck groups.
Any topological spherical three-manifold appears on its universal cover $S^3$ in the form of a~tiling. 
Any tile is an~image of the spherical manifold. 
The tiles are related to one another by deck actions from the deck group. 
We construct from the homotopies of~\cite{EV04} the isomorphic deck actions on the three-sphere $S^3$ and, from them 
determine, the deck groups for the Platonic manifolds.
The deck groups are factorized in Section~\ref{sec:point} by the point symmetry groups of the Platonic polyhedra. 
We handle fixpoints under rotations in point groups by the new construction of multiply-connected topological orbifolds. 
Whereas a~topological manifold strictly excludes fixpoints under deck actions, orbifolds allow for such fixpoints
\cite{TH97,MO87}, \cite{RA94}~pp.~652--714.

For the harmonic analysis on orbifolds, we employ in Section~\ref{sec:orbi} Wigner polynomials with restrictions to the subgrid representation of topological selection rules.
The resulting subbases appear as denumerable sets of Wigner polynomials on subgrids, as shown in Figure~\ref{fig:subgrids} and extend earlier work~\cite{KR10a}.
The $m$-grid representation allows in Section~\ref{sec:harmonic} for a~very transparent graphical display of 
the restriction from general bases to subbases of a given topology.

Finally in Section~\ref{sec:cmb}, we present the application of topology to cosmology.
In contrast to Einstein's closure of the spatial part of space-time on the simply-connected three-sphere,
we discuss the closure on spherical orbifolds and incorporate a~wider variety of multiply-connected cosmologies. 
Our harmonic analysis on orbifolds yields for each orbifold specific multipole selection rules and, moreover, predicts topological correlations between different multipole orders $(l,l')$ of the CMB.

\section{The Three-Sphere $S^3$ Is Unitary}\label{sec:3sphere}

In three-dimensional topology, the three-sphere $S^3$ is the simply-connected universal cover~\cite{TH97} (p.~290) of~spherical topologies. 
The points of the three-sphere, seen in Euclidean space $E^4$ with covariant coordinates, are: 
\begin{equation}
 \label{E2} x=(x_0,x_1,x_2,x_3): \sum_{i=0}^3 (x_i)^2=1
\end{equation}
They are in one-to-one correspondence to the elements of the unitary unimodular group $SU(2,C)$. 
From real coordinates of $E^4$: $x=(x_0,x_1,x_2,x_3)$, we pass to a~$2 \times 2$ unimodular unitary matrix $u(x)\in SU(2,C)$ in the form:
\begin{equation}
\label{u1} 
u(x)=\left[
\begin{array}{ll}
 z_1&z_2\\
-\overline{z}_2&\overline{z}_1\\
\end{array}
\right]
= \left[
\begin{array}{ll}
 x_0-ix_3&x_2-ix_1\\
-x_2-ix_1&x_0+ix_3\\
\end{array}
\right]
, \: {\rm det} (u)=1
\end{equation} 
As angular coordinates on $SU(2,C)$, we can use the Euler angles $(\upalpha,\upbeta,\upgamma)$~\cite{ED57} (pp.~6--8). 
In these, the matrix $u$ takes the form:
\begin{eqnarray}
\label{u1b} 
&u(\upalpha,\upbeta,\upgamma)= \left[ \begin{array}{ll}
\exp(\frac{i(\upalpha+\upgamma)}{2}) \cos(\frac{\upbeta}{2})&\exp(\frac{i(\upalpha-\upgamma)}{2}) \sin(\frac{\upbeta}{2})\\
  -\exp(\frac{-i(\upalpha-\upgamma)}{2}) \sin(\frac{\upbeta}{2})&\exp(\frac{-i(\upalpha+\upgamma)}{2}) \cos(\frac{\upbeta}{2})\\ 
  \end{array} \right]
\end{eqnarray}
The measure of integration in the Euler angles becomes~\cite{ED57} (p.~62),
\begin{equation}
 \label{u1d} 
d\upmu(\upalpha,\upbeta,\upgamma)= \frac{1}{8\pi^2} d\upalpha\sin(\upbeta)d\upbeta d\upgamma 
\end{equation}

{Summary:} The three-sphere $S^3$ corresponds one-to-one to the group $SU(2,C)$. From this correspondence, the isometries of $S^3$ inherit a~unitary structure. 
This provides the full representation theory~\cite{ED57} of the group $SU(2,C)$ 
as a tool for spherical topology.

\section{Wigner Polynomials}\label{sec:wigner}

The finite irreducible unitary representations of $SU(2,C)$ were studied by Wigner~\cite{WI59}.
For~unitarity and the irreducibility of representations, we refer to his monograph.
We now introduce the Wigner polynomials. We shall show that they span the harmonic analysis on the three-sphere~\cite{KR05} and play an~important part in topological analysis.

Let the matrix $u\in SU(2,C)$ Equation~(\ref{u1}) act from the left on the complex column vector:
\begin{equation}
\label{u5}
u: \left[\begin{array}{l}
 a\\
 b
 \end{array}\right]\\
\rightarrow u(z_1,z_2)\left[\begin{array}{l}
 a\\
 b
 \end{array}\right]
\end{equation}
In the Bargmann--Hilbert space of analytic functions~\cite{BA68} in two complex variables $(a,b)$, the monomials:
\begin{equation}
 \label{u6} 
\upphi(a,b)^j_m = \frac{1}{\sqrt{(j-m)!(j+m)!}} a^{j+m}b^{j-m}, 2j=0,1,2,.., -j, \leq m\leq j 
\end{equation}
are normalized with the measure:
\begin{equation}
\label{u7}
{\rm d}\upmu(a,b)= \pi^{-2} \exp(-a\overline{a}-b\overline{b})\: {\rm d}\text{Re}(a){\rm d}\text{Im}(a)\: {\rm d}\text{Re}(b){\rm d}\text{Im}(b)
\end{equation}
With $m$ either an integer or half-integer, under the action Equation~(\ref{u5}), the monomials in Equation~(\ref{u6}) carry irreducible representations of $SU(2,C)$ and transform as: 
\begin{equation}
\label{u8}
\upphi(az_1+bz_2,-a\overline{z}_2+b\overline{z}_1)^j_{m_1}
=\sum_{m_2}\upphi(a,b)^j_{m_2} D^j_{m_2,m_1} (z_1,z_2,\overline{z}_1,\overline{z}_2)
\end{equation}
The $(2j+1)^2$ coefficients in this equation are the Wigner $D^j$-functions~\cite{ED57} (Equation~(4.1.10)). 
Wigner~\cite{WI59} has shown that the $D^j$-functions are the unitary irreducible representations of
$SU(2,C)$. 
We replace Euler angles by four complex variables from Equation~(\ref{u1}) in the arguments of the $D^j$.
Following~\cite{KR05}, we term the resulting complex valued polynomials Wigner polynomials. 
From Equation~(\ref{u8}), they are given~by:
\begin{eqnarray}
\label{u8a}
D^j_{m_1,m_2}(z_1,z_2,\overline{z}_1,\overline{z}_2):&=&
\left[\frac{(j+m_1)!(j-m_1)!}{(j+m_2)!(j-m_2)!}\right]^{1/2}\nonumber\\\nonumber
&& \sum_{\upsigma}
\frac{(j+m_2)!(j-m_2)!}{(j+m_1-\upsigma)!(m_2-m_1+\upsigma)!\upsigma!(j-m_2-\upsigma)!}
\\ 
&& \times (-1)^{m_2-m_1+\upsigma} z_1^{j+m_1-\upsigma}\overline{z}_2^{m_2-m_1+\upsigma}z_2^{\upsigma} 
\overline{z}_1^{j-m_2-\upsigma} \vspace{-10pt}
\label{u9}
\end{eqnarray}
All exponents of the four complex variables in Equation~(\ref{u8a}) are integers and must be non-negative. 
This restricts the summation over the integer $\upsigma$.
Two important properties of these Wigner polynomials arise under inversion and under complex conjugation of the matrix 
$u$,
\begin{equation}
 \label{u10}
D^j_{m_1m_2}(u^{-1})= \overline{D^j_{m_2m_1}(u)},\: D^j_{m_1m_2}(\overline{u})= \overline{D^j_{m_1m_2}(u)}
\end{equation}
We prove that the Wigner polynomials vanish under the Laplacian on $E^4$ and, so, are harmonic. 
Consider first the case $m_1=j$. This implies $\upsigma=j-m_2$,
\begin{equation}
 \label{u11}
 D^j_{j,m_2}(u) = \left[\frac{(2j)!}{(j+m_2)!(j-m_2)!}\right]^{1/2} (z_1)^{j+m_2}(z_2)^{j-m_2}
\end{equation}
The polynomials Equation~(\ref{u11}) are analytic in $(z_1,z_2)$, and so, by the Cauchy--Riemann differential equations, vanish under the Laplacian $\Updelta$ on $E^4$. 
From the particular polynomials Equation~(\ref{u11}), 
we can lower the eigenvalue $m_2=j$ to any value $m_2=j-1,.., -l$ for fixed degree $2j$ by the application of the left lowering operator $L^l_-$ of ${SU(2,C)}^C$, given from~\cite{KR05} (Equation~(35)) by: \vspace{-4pt}
\begin{equation}
\label{u11b}
L^l_-=[\overline{z}_2\partial_{z_1}-\overline{z}_1\partial_{z_2}] \vspace{-6pt}
\end{equation}
By repeated application of this lowering operator, we can reach all of the $(2j+1)^2$ Wigner polynomials for fixed degree $2j$.
The lowering operator commutes with the Laplacian $\Updelta$ on $E^4$ and, so, cannot change its eigenvalue zero. 
It follows that all of the Wigner polynomials Equation~(\ref{u8a}) are harmonic.

The action of the rotation group $SO(4,R)$ is expressed by left and right actions on $SU(2,C)$. 
When restricted to conjugation, its action reduces to rotations only of the three coordinates 
 $(x_1,x_2,x_3)\in~E^3$; see Equation~(\ref{u14db}).
The orthogonality of the Wigner polynomials, expressed in the Euler angles, \linebreak is~\cite{ED57} (p. 62)
\begin{eqnarray}
 \label{u11a} 
&&\frac{1}{8\pi^2}\int \overline{D^{j'}_{m_1'm_2'}(\upalpha \upbeta \upgamma)} D^j_{m_1m_2}(\upalpha\upbeta\upgamma)
d\upalpha\sin(\upbeta)d\upbeta d\upgamma
\\ \nonumber
&&= \updelta_{j', j}\updelta_{m_1', m_1}\updelta_{m_2', m_2}\frac{1}{2j+1}
\end{eqnarray}

{Summary:} The Wigner polynomials, Equation~(\ref{u6}), are harmonic, orthonormal and homogeneous of degree $2j=0,1,2,...,\infty$. Their transformation properties under $SU(2,C)$ are well known from the theory of angular momentum in quantum mechanics 
\cite{ED57}.

\section{Unitary Actions and Representations}\label{sec:unitary}

The group of isometries of $S^3$ is $SO(4,R)$. 
The action of this group is isomorphic to the direct product of a~left and a~right group $SU^l(2,C), SU^r(2,C)$ acting on $u \in SU(2,C)$ in the form:
\begin{equation}
\label{u2}
SO(4,R) \sim (SU^l(2,C)\times SU^r(2,C))/Z_2
\end{equation}
The left and right action groups in Equation~(\ref{u2}) commute and act as: 
\begin{equation}
\label{u3}
(g_l,g_r)\in (SU^l(2,C)\times SU^r(2,C)): u \rightarrow g_l^{-1}ug_r
\end{equation}
The subgroup $Z_2$ in Equation~(\ref{u2}) is generated by $(g_l, g_r)=(-e,-e)\in (SU^l(2,C)\times SU^r(2,C))$.
The direct product form Equation~(\ref{u2}), contrary to what is asserted in~\cite{LE02} (p.~277), in general does not extend to
subgroups of $SO(4,R)$ and, so, cannot yield their classification. 
A counter-example is provided by the deck group $C_5$ of the tetrahedral manifold (Equation~(\ref{f12})), which entangles 
left and right actions.

Among the actions described by Equation~(\ref{u3}) are the conjugation actions: \vspace{-3pt}
\begin{equation}
\label{u3b}
(g_l, g_l): u \rightarrow g_l^{-1}u g_l \vspace{-3pt}
\end{equation}
which form 
a subgroup $SU(2,C)^C$. The action Equation~(\ref{u3b}) preserves the trace, \vspace{-3pt}
\begin{equation}
 \label{u3c} 
{\rm trace}(u(x))= 2x_0 \vspace{-3pt}
\end{equation}
and therefore operates only on the 
subspace with coordinates $(x_1,x_2,x_3) \in E^3$.

The unitary structure of the three-sphere governs its isometries and 
introduces the representation theory of~$SU(2,C)$.

In particular, we obtain for the action of $SO(4,R)$ on a~Wigner polynomial Equation~(\ref{u9}): 
\begin{eqnarray}
 \label{u14}
 (T_{(g_l,g_r)}D^j_{m_1,m_2})(u) &=& D^j_{m_1,m_2}(g_l^{-1}ug_r) = \sum_{m_1'm_2'} D^j_{m_1'm_2'}(u)\: D^{(j,j)}_{(m_1'm_2',m_1 m_2)}(g_l,g_r)\\
 D^{(j,j)}_{(m_1'm_2',m_1 m_2)}(g_l,g_r)
 &=&D^j_{m_1m_1'}(g_l^{-1})D^j_{m_2'm_2}(g_r) \vspace{-2pt}
\end{eqnarray}
The expression in the last line is the matrix element of the irreducible representation $D^{(j,j)}(g_l,g_r)$ of $SO(4,R)$
for the pair $(g_l,g_r) \in SU^l(2,C)\times SU^r(2,C)$. 
The degree $2j$ stays fixed under the action of $SO(4,R)$, while the pairs $(m_1,m_2)$ take $(2j+1)^2$ values.

The actions by conjugation Equation~(\ref{u3b}) of the subgroup ${SU(2,C)}^C$ reduce into irreducible form~upon transforming the Wigner polynomial basis; compare also~\cite{AU08}, with Wigner coefficients of $SU(2,C)$~\cite{ED57} (pp.~31--52), into a~spherical basis by: 
\begin{eqnarray}
 \label{u14b}
\uppsi_{\upbeta lm}(u) & = &\updelta_{\upbeta,2j+1}\sum_{m_1m_2} D^j_{m_1,m_2}(u) C(jm_1m_2|lm)
\\ \nonumber 
C(jm_1m_2|lm)&:=&\langle j-m_1jm_2|lm\rangle (-1)^{(j-m_1)}, \quad l=0,1,\ldots,2j, \quad 2j=\upbeta-1
\end{eqnarray}
Equation~(\ref{u14b}) implies that the irreducible representation $D^{(j,j)}$ of $SU(4,R)$ when reduced to the subgroup $SU(2,C)^C$ contains 
any irreducible representation $D^l$ of $SU(2,C)^C$ once and only once for $l=0,\ldots,2j$. 
The transformation inverse to Equation~(\ref{u14b}) is: 
\begin{equation}
 \label{u14c}
D^j_{m_1,m_2}(u)= \updelta_{\upbeta, 2j+1}\updelta_{m,-m_1+m_2} 
\sum_{l=0}^{2j}\uppsi_{\upbeta lm}(u) C(jm_1m_2|lm),\quad l\in Z,\: -l\leq m\leq l
 \end{equation}
The two transformations Equations~(\ref{u14b}) and (\ref{u14c}) will play a~crucial role in the recursive construction of
the harmonic analysis for orbifolds in Section~\ref{sec:projharmonic}. 
In cosmic topology, Section~\ref{sec:cmb}, the spherical basis Equation~(\ref{u14b}) in terms of spherical harmonics $Y_{m}^l$ is required for the multipole analysis of physical observables, like the CMB radiation.
It is given by:
\begin{eqnarray}
\uppsi_{\upbeta lm}(u)&=&R_{\upbeta l}(\upchi) Y_{m}^l(\uptheta,\upphi)\\
R_{\upbeta l}(\upchi)&=&2^{l+1/2} l! \sqrt{\frac{\upbeta(\upbeta-l-1)!}{\pi(\upbeta+l)}} C^{l+1}_{\upbeta-l-1}(\cos(\upchi))
\end{eqnarray}
where $C^{l+1}_{\upbeta-l-1}$ denotes the Gegenbauer polynomial.

As a~result of Equation~(\ref{u14b}), the conjugation action of $SU(2,C)^C$ on the spherical basis becomes: \vspace{-3pt}
\begin{eqnarray}
\label{u14db}
(T_{(g,g)}\uppsi)_{\upbeta lm}(u) = \uppsi_{\upbeta lm}(g^{-1}ug)= \sum_{m'=-l}^l \uppsi_{\upbeta lm'}D^l_{m',m}(g) \vspace{-6pt}
\end{eqnarray}
identical to the irreducible action of the rotation operator $T_g$ on the usual spherical harmonics $Y^l_m$ in $E^3$~\cite{ED57} (pp.~53--67). The coordinate $x_0$ 
in $E^4$ is unchanged under this action, and so, we find:%please check the correction

{Summary:} The action of the rotation $(g_l,g_r)\in SO(4,R)$ on Wigner polynomials is given by the representation 
$D^{(j,j)}(g_l,g_r)$, Equation~(\ref{u14}). The rotations $(g,g) \in SO(3,R)$ act only on the 
first three~coordinates, and the spherical basis Equation~(\ref{u14b}) yields the decomposition of Wigner polynomials into irreducible subbases 
characterized by the multipole order $l$.

\section{Discrete $2D$ $m$-Grids}\label{sec:grids}

To display the topological selection rules for Wigner polynomials, we concentrate on their discrete labels $(m_1, m_2)$.
Both of them are an integer or half-integer. 
These labels form an~integer plus a~half-integer $2D$ $m$-grid, both of spacing $(\pm 1,\pm 1)$, on a~$2D$ plane. From the point of view of representations of $SU(2,C)$,
the grid points fix subrepresentations of left and right subgroups $U(1)$.
Consider the set of Wigner polynomials as being attached to the grid points $(m_1,m_2)$. 
At a~given grid point $(m_1,m_2)$, the labels $j$ of all Wigner polynomials attached to it have the denumerable range $j= j_0, j_0+1,\ldots, \infty,\:j_0= {\rm Max}(|m_1|,|m_2|)$. 
Conversely, the Wigner polynomials for fixed degree $2j$ occur with values $(m_1, m_2)$ on a~centered square $(|m_1|,|m_2| \leq j)$.
In Section~\ref{sec:subgrids}, we shall see that harmonic analysis on topological orbifolds selects Wigner bases on subgrids.

{Summary:} A~denumerable set of Wigner polynomials $D^j(u)$ with the index pair $(m_1,m_2)$ and degree $2j, j \geq (|m_1|,|m_2|)$ is associated with any single point $(m_1,m_2)$ of a~$2D$ $m$-grid. 

\section{Spherical Coxeter Groups for the Platonic Polyhedra}\label{sec:coxeter}

\begin{figure}[tb]
\begin{center}
\includegraphics[width=0.8\textwidth]{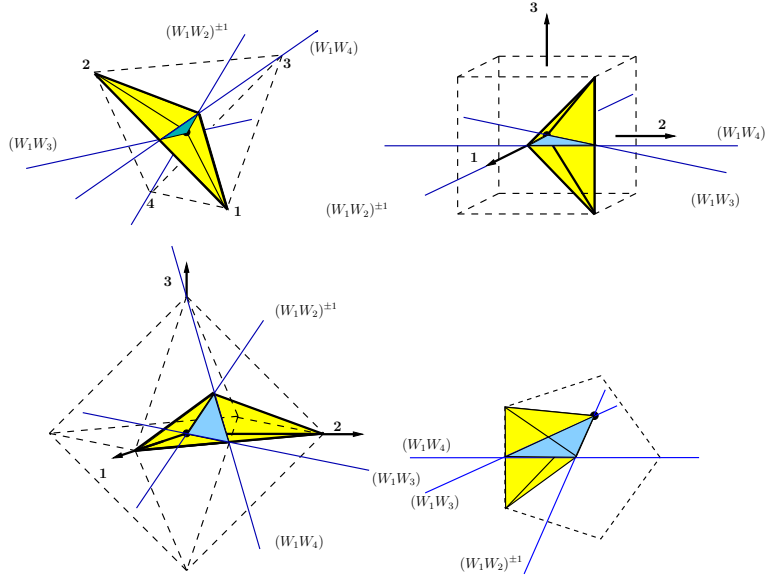} 
\end{center}
\caption{\label{fig:duplices}
Geometric orbifolds $N1$, $N2$ are shown in the top panels and $N3$, $N9$ in the bottom panels. 
The yellow color marks the duplex form of the orbifolds as pairs of Coxeter simplices, glued with 
triangular faces in blue color. 
The axes of the covering rotations $(W_a W_b)$ for each orbifold are marked by thin lines. 
The parts of these axes that bound the glue triangles carry fixpoints of integer order $p$.
}
\end{figure}

In this section, we turn from the three-sphere to the spherical manifolds of Platonic polyhedra shown in Figure~\ref{fig:duplices}.
To describe their geometrical transformation properties, we use spherical Coxeter groups.
Coxeter groups $\Upgamma$ are generated by Weyl reflections in (hyper-)planes of Euclidean space $E^n$. Any Weyl reflection $W_a$ is characterized by a~unit vector $a$, normal to the reflection (hyper-)plane. The action of the Weyl reflection $W_a$ on $x \in E^n$ is given by: \vspace{-4pt}
\begin{equation}
 \label{u15}
 W_a: x \rightarrow W_ax= x-2\langle x, a~\rangle,\: (W_a)^2=I \vspace{-4pt}
\end{equation}
A Weyl reflection is a~Euclidean isometry of determinant $-1$. 
A Coxeter group $\Upgamma$ is generated by Weyl reflections $W_{a_i}, i=1,2,\ldots$. 
For a~fixed Coxeter group $\Upgamma$, we use the short-hand notation $W_{a_1}= W_i$. 
$\Upgamma$ is graphically denoted by a~Coxeter diagram, built from circles for each generator and connected by a~sequence of lines. The diagram allows one to construct the group $\Upgamma$ from its Weyl reflection generators and 
their relations~\cite{MA66} (pp.~12--23),~\cite{HU90}. The lines between circles carry integer numbers $m_{ij}$. Lines without numbers are short-hand for $m_{ij}=3$. The number $m_{i,i+1}$ between two successive reflections $W_{a_i},W_{a_{i+1}}$ implies the relation: \vspace{3pt}
\begin{equation}
\label{u16}
(W_{a_i}W_{a_{i+1}})^{m_{i,i+1}}= I \vspace{2pt}
\end{equation}
The number $m_{i,i+1}$ is the order of the rotation generated by $W_{a_i}W_{a_{i+1}}$.

In Tables~\ref{tab:coxa} and \ref{tab:coxb}, we list the four spherical Coxeter groups of the tetrahedron, cube, octahedron and dodecahedron, each with four generators, their diagrams and quadruples of unit vectors $a_1,\ldots,a_4$. 
These Coxeter groups produce tilings~\cite{SO58} of the three-sphere $S^3$ by 
Platonic polyhedra. 

\begin{table}[H]
\small
\centering
\begin{tabular}{llllll}\toprule
{\bf Coxeter Diagram} $\Upgamma$ & $|\Upgamma|$ & {\bf Polyhedron} ${\cal M}$ & $H={\rm deck}({\cal M})$ & $|H|$ & {\bf Reference} \\ \midrule
$\circ -\circ -\circ - \circ$   & $120$ & tetrahedron $N1$ & $C_5$  & $5$ &~\cite{KR08} \\ \midrule
$\circ \stackrel{4}{-} \circ -\circ -\circ$ & $384$ & cube $N2$ & $C_8$  & $8$ &~\cite{KR09} \\                                     
         & & cube $N3$ & $Q$  & $8$ & ~\cite{KR09} \\ \midrule       
$\circ -\circ \stackrel{4}{-}\circ - \circ$ & $1152$ & octahedron $N4$ & $C_3\times Q$ & $24$&\cite{KR10a} \\
         & & octahedron $N5$ & $B$  & $24$&\cite{KR10a} \\
         & & octahedron $N6$ & ${\cal T}_2$ & $24$&\cite{KR10a} \\ \midrule
$\circ -\circ -\circ \stackrel{5}{-} \circ$ & $120\cdot 120$ & dodecahedron $N9$ & ${\cal J}_2$ & $120$ &~\cite{KR05} 
\\ \bottomrule
\end{tabular}
\caption{\label{tab:coxa}
Diagrams of four spherical Coxeter groups $\Upgamma$ of order $|\Upgamma|$, four Platonic polyhedra ${\cal M}$ and seven deck groups $H={\rm deck}({\cal M})$ of order $|H|$ according to~\cite{CM65} (p.~134).
In the table, $C_n$ denotes a~cyclic, $Q$ the quaternion, ${\cal T}_2$ the binary tetrahedral and ${\cal J}_2$ the binary icosahedral group. The symbols $Ni$ are adapted from~\cite{EV04}.
}
\end{table}
\vspace{-6pt}
\begin{table}[H] \vspace{-12pt}
\centering
\begin{equation*}
\begin{array}{lllll} 
\toprule
{\bf \Upgamma} & {\bf a_1} & {\bf a_2} & {\bf a_3}& {\bf a_4}\\ \midrule
\circ -\circ -\circ - \circ & (0,0,0,1)&(0,0,\sqrt{\frac{3}{4}},\frac{1}{2})
& (0,\sqrt{\frac{2}{3}},\sqrt{\frac{1}{3}},0)& (\sqrt{\frac{5}{8}},\sqrt{\frac{3}{8}},0,0)\\ \midrule
\circ \stackrel{4}{-} \circ -\circ -\circ& (0,0,0,1)& (0,0,-\sqrt{\frac{1}{2}},\sqrt{\frac{1}{2}})
&(0,\sqrt{\frac{1}{2}},-\sqrt{\frac{1}{2}},0)&(-\sqrt{\frac{1}{2}},\sqrt{\frac{1}{2}},0,0)\\ \midrule
\circ -\circ \stackrel{4}{-}\circ - \circ& (0,\sqrt{\frac{1}{2}},-\sqrt{\frac{1}{2}},0)&(0,0,-\sqrt{\frac{1}{2}},\sqrt{\frac{1}{2}})
&(0,0,0,1)& (\frac{1}{2},\frac{1}{2},\frac{1}{2},\frac{1}{2})\\ \midrule
\circ -\circ -\circ \stackrel{5}{-} \circ& (0,0,1,0)&(0,-\frac{\sqrt{-\uptau+3}}{2},\frac{\uptau}{2},0)
&(0,-\sqrt{\frac{\uptau+2}{5}},0,-\sqrt{\frac{-\uptau+3}{5}})&(\frac{\sqrt{2-\uptau}}{2},0,0,-\frac{\sqrt{\uptau+2}}{2}) \\ \bottomrule
\end{array}
\end{equation*}
\caption{\label{tab:coxb}
The four Weyl vectors $a_s, s=1,\ldots,4 \in E^4$ 
for the four Coxeter groups $\Upgamma$ listed in Table~\ref{tab:coxa}, with $\tau:=\frac{1+\sqrt{5}}{2}$.
} 
\end{table}
\vspace{-6pt}
A spherical Coxeter group possesses a~fundamental domain. On $E^4$, this is a~simplex bounded by the four Weyl reflection planes from the generators of $\Upgamma$. 
For the Platonic manifolds, duplex pairs of them are shown in Figure~\ref{fig:duplices}.

{Summary:} Each Platonic spherical polyhedron is linked to one of the four Coxeter groups listed in Tables~\ref{tab:coxa} and \ref{tab:coxb}.

\section{The Unimodular Subgroups $S\Upgamma$}\label{sec:unimodular}

Any element of a~Coxeter group from Tables~\ref{tab:coxa} and \ref{tab:coxb} containing an~even number of reflections is a~rotation. 
For topology, we need these rotations, since they preserve orientation. Their $4 \times 4$ representations on $E^4$ are unimodular with determinant 
one. The subgroup of a~Coxeter group $\Upgamma$ generated by all rotations has unimodular matrix representations, and we denote it by $S\Upgamma$. This subgroup for all four Coxeter groups in Table~\ref{tab:coxa} is 
generated in each case by the three products $((W_1 W_2), (W_2W_3), (W_3W_4))$ of Weyl reflections. 

We wish to pass from a~product $(W_a W_b)$ of two Weyl reflection operators to the standard form Equation~(\ref{u14}) of $SO(4,R)$
with two $SU(2,C)$ group parameters $(g_l,g_r)$. 
First, we determine from the unit reflection vectors $a,b$ two unitary matrices $v(a), v(b)$ by use of 
Equation~(\ref{u1}),
\begin{equation}
\label{b2}
v(a) :=
\left[\begin{array}{ll}
 a_0-ia_3&a_2-ia_1\\
 -a_2-ia_1& a_0+ia_3
 \end{array}\right]
\end{equation} 
and similarly for $v(b)$.
Then, as shown in~\cite{KR10} (Equation~(11)), we find between products of Weyl operators and 
$SO(4,R)$ actions Equation~(\ref{u14}) the operator relation: 
\begin{equation}
 \label{b3}
T_{W_a W_b} \equiv T_{(g_l,g_r)}: (g_l,g_r)=(v(a)v^{-1}(b), v^{-1}(a)v(b)) 
\end{equation}

{Summary:} The unimodular subgroup $S\Upgamma$ of a~Coxeter group $\Upgamma$ is generated by products $(W_a W_b), \linebreak a~\neq b$ of pairs of Weyl reflection operators. Any such product can be converted by Equations~(\ref{b2}) and~(\ref{b3}) into
a rotation operator with parameters $(g_l,g_r)$.

\section{Homotopy with Spherical Polyhedra}\label{sec:homotopy}

There are five spherical Platonic polyhedra.
Everitt in~\cite{EV04} applied Sim's low index subgroups algorithm to determine all possible homotopies of these polyhedra. 
His results are given in the form of diagrams for numbered faces and edges and identification of pairs of faces and edges. 
The identification is also referred to as gluings and by construction encodes the first homotopy or fundamental group.

\begin{figure}[t]
\begin{center}
% Figure 2 is central for the argument in this article and contains important labels
% please maintain readability and do not shrink it below 0.7\textwidth 
% Thanks! PK
\includegraphics[width=0.7\textwidth]{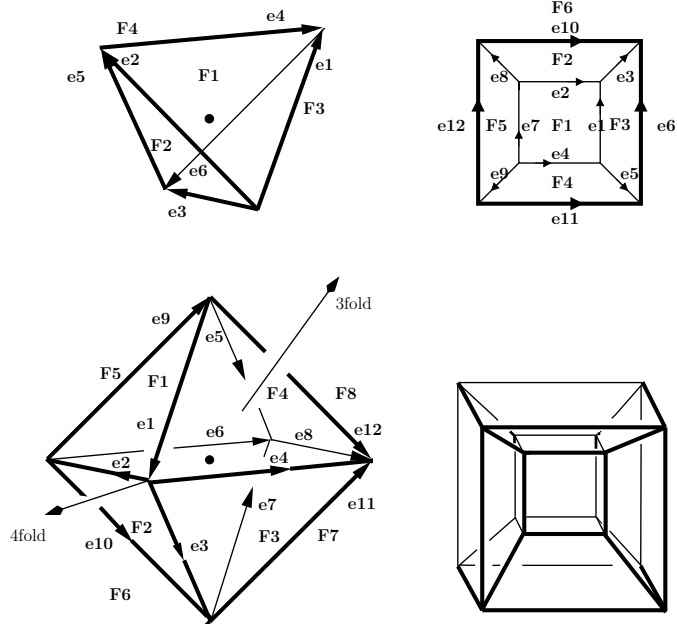} 
\end{center}
\caption{\label{fig:platonicpoly}
Face and edge enumeration of the tetrahedron, cube, octahedron 
and the eight-cell that divides the three-sphere into eight spherical cubes.} 
\end{figure} 

We prefer to list the gluings of faces and edges, enumerated in line 
with Everitt~\cite{EV04} and shown in Figure~\ref{fig:platonicpoly}. Note that a~given spherical polyhedron may have several different inequivalent homotopy groups.
As standard examples, we take the cubic spherical manifolds~\cite{KR09}. 
This manifold admits two~inequivalent homotopies, which we denote as $N2,N3$. 
We depict the polyhedra in an~Euclidean fashion. For the cube, an~enumeration of the six faces $Fi, i=1,...,6$ and twelve directed edges $1,...,12$ following Everitt~\cite{EV04} is shown in Figure~\ref{fig:platonicpoly}. Any square face is characterized by a~directed sequence of four edges. Any homotopic gluings of faces plus edges is a~map between two such sequences. As an example, we illustrate a~gluing $F1\cup F3$ in $N2$ as a~map from right to left, 
\begin{equation}
\label{u51}
g_1(1 \Leftarrow 3)
= \left[\begin{array}{lll}
 &\overline{1}& \\
 \overline{4}&&2\\
& 7& 
\end{array}\right]
\Leftarrow 
\left[\begin{array}{lll}
&\overline{3}&\\
 \overline{1}&&6 \\
 & 5& 
 \end{array}\right]
 \end{equation}

From~\cite{EV04}, we find for the first cubic homotopy group $N2$ the pairwise face gluings: \vspace{-3pt}
\begin{equation}
 \label{u52} 
F3 \cup F1, F4\cup F2, F6 \cup F5 \vspace{-6pt}
\end{equation}
The glued directed edges are listed on single lines in the diagram given in Table~\ref{tab:homo}.

{Summary:} All possible homotopies for the Platonic spherical polyhedra are determined and listed 
in Table~\ref{tab:homo} in accord with~\cite{EV04}.

\begin{table}[t]
\footnotesize
\centering
\begin{tabular}{lll} \toprule
{{\bf Polyhedron}}& {\rm {\bf Face}\: {\bf Glue}}& {\rm {\bf Edge}\: {\bf Glue}} \\ \midrule

{Tetrahedron} $N1$& $F3 \cup F1,\: F2 \cup F4$   & $\left[ \begin{array}{lll}
             1&\overline{3}&\overline{4}\\
             2&\overline{5}&\overline{6}\\
             \end{array}\right]$ \\ \midrule
            
{Cube} $N2$ & $F3 \cup F1,\: F4\cup F2,\: F6\cup F3$ & $\left[ \begin{array}{lll}
             1&3&4\\
             2&6&\overline{9}\\
             5&7&\overline{10}\\
             8&11&\overline{12}\\
             \end{array}\right]$\\ \midrule
               
{Cube } $N3$ & $F1 \cup F6,F2 \cup F4, F3 \cup F5$ & $\left[ \begin{array}{lll}
             1&8&11\\
             2&\overline{6}&\overline{9}\\
             3&4&\overline{12}\\
             5&\overline{7}&\overline{10}\\
             \end{array}\right]$\\ \midrule 
               
{Octahedron} $N4$ & $F6 \cup F2, F5 \cup F3, F1 \cup F4, F7 \cup F8$ & $\left[ \begin{array}{lll}
             1&4&9\\
             2&7&\overline{12}\\
             3&6&\overline{10}\\
             5&8&11\\
             \end{array}\right]$\\ \midrule
               
{Octahedron} $N5$ & $F6 \cup F8, F1 \cup F4, F2 \cup F7, F3 \cup F5$ & $\left[ \begin{array}{lll}
             1&4&9\\
             2&\overline{7}&\overline{12}\\
             3&6&8\\
             5&\overline{10}&11\\
             \end{array}\right]$\\                     
             \midrule           
{Octahedron} $N6$ & $F6 \cup F4, F5 \cup F3, F8 \cup F2, F7 \cup F1$ & $ \left[ \begin{array}{lll}
             1&8&10\\
             2&5&11\\
             3&6&12\\
             4&7&9\\
             \end{array} \right]$\\
             \bottomrule
\end{tabular} 
\caption{\label{tab:homo} The homotopies of Platonic spherical polyhedra in terms of face and edge gluings. The directed edges in each line of the last column are glued.}
\end{table}

\section{Synopsis of Platonic Homotopies}\label{sec:synopsis}

The enumeration of faces and directed edges of the Platonic polyhedra is shown in Figure~\ref{fig:platonicpoly}.
In Table~\ref{tab:homo}, we list the gluing of faces and directed edges as given in~\cite{EV04}. 

\section{From Homotopies to Deck Actions on $S^3$}\label{sec:deck}

Homotopic gluings relate faces and edges of a~single polyhedron. A~new geometric view of topologies emerges on the universal covering manifold $S^3$. A~general theorem, given by Seifert and Threlfall~\cite{SE34} (pp.~181--203), proves for topological manifolds the isomorphism between the first homotopy group and the group of deck or covering actions on the universal cover.

We implement this theorem for the cubic spherical manifold $N2$ and convert its homotopic gluings from Table \ref{tab:homo} into deck actions between neighboring copies of a~single proto-cube with fixed face and edge enumeration. The deck rotations, which generate the tiling of $S^3$, form the topological deck group. The eight cubic copies tile the three-sphere in the form of the eight-cell shown in Figure~\ref{fig:platonicpoly} in a~projection from~\cite{SO58} (p.~170). This means that the topological deck group must be of order eight.

For the homotopy of $N2$, we make full use of the prescriptions of Table \ref{tab:homo}.
Any quadratic face of the proto-cube is surrounded by four directed edges given in Figure~\ref{fig:platonicpoly}. 
These four directed edges in turn determine the face and, moreover, give its orientation with respect to a~four-fold axis perpendicular to the~face.

We use square diagrams as given in Equation~(\ref{u54}), formed from four directed edges, to denote edges around a~face plus their orientation.
This orientation must be respected in the homotopic gluing of faces and their edges given in Table \ref{tab:homo}.

It suffices to convert only the homotopic face gluings of a~proto-cube, the glue generators, into deck actions. Then, these 
deck actions by multiplication generate the full deck group.

In $E^3$, we choose the coordinates $(x_1,x_2,x_3)$ normal to the three faces labeled $F1, F3, F2$ in Figure~\ref{fig:platonicpoly}. Positive rotations by an~angle $\upphi$ around these normals we denote as 
$R_i(\upphi), i=1,2,3$.

Before turning to the manifold $N2$ in Equation~(\ref{u55}), we first treat in Equation~(\ref{u54}) 
a simple reference glue $F1\cup F6$ of two opposite faces of the proto-cube. We convert this glue into a~deck action of two~copies of the proto-cube and write it algebraically in Equation~(\ref{u54}) as a~product of Weyl reflections and inversions. By ${\cal J}_3, {\cal J}_4$, 
we denote the inversions 
in $E^3, E^4$. We find this action in the form: \vspace{4pt}
\begin{eqnarray}
 \label{u54}
&& st(1 \Leftarrow 6)
= \left[\begin{array}{lll}
 &\overline{4}& \\
 7&&\overline{1}\\
&2& 
\end{array}\right]
\Leftarrow 
\left[\begin{array}{lll}
&\overline{10}&\\
 \overline{6}&&12 \\
 &\overline{11}& 
 \end{array}\right] 
\\ \nonumber 
&& = W_4 {\cal J}_3= W_4 W_0 {\cal J}_4
\end{eqnarray}
The product of Weyl reflections and inversions in Equation~(\ref{u54}) combines into an~overall unimodular rotation matrix. For convenience, we introduced the additional Weyl reflection operator $W_0$, \linebreak $a_0=(0,0,0,1)$.

Now, we return to the cubic manifold $N2$. By use of two rotations, we convert its first face gluing, $st(1 \Leftarrow 3): F1 \cup F3$, into the reference deck action Equation~(\ref{u54}), \vspace{-4pt}
\begin{equation}
 \label{u55}
 st(1 \Leftarrow 3)= R_1(\pi/2) st(1 \Leftarrow 6) R_3(\pi/2) \vspace{-4.5pt}
\end{equation}
By similar conjugations as in Equation~(\ref{u54}), we convert the two other face gluings of the manifold $N2$ in Equation~(\ref{u52}) into deck actions, \vspace{-6pt}
\begin{eqnarray}
\label{u56} 
&&st(2 \Leftarrow 4)= (W_3W_2)st(1 \Leftarrow 6)(W_2W_3)
\\ \nonumber 
&&st(3 \Leftarrow 5)= (W_2W_3)st(1 \Leftarrow 6)(W_3W_2) 
\end{eqnarray} 
From products of two Weyl reflection operators in Equation~(\ref{u56}), we pass to elements $(g_l,g_r) \in SO(4,R)$ given in Table~\ref{tab:coxb} by Equation~(\ref{b3}).
As shown in~\cite{KR10}, the deck generator in Equation~(\ref{u55}) generates a~cyclic group $C_8$ of order eight, which paves 
the eight-cell tiling of $S^3$ by copies of the~prototile. 

The eight-cell admits a~second inequivalent cubic homotopy $N3$~\cite{EV04}. 
Its deck group derived in~\cite{KR10} is the quaternion group $Q$. 
We give the generating elements in Equation~(\ref{qg}).

For the other Platonic homotopies studied by Everitt~\cite{EV04}, the conversions to deck actions are carried out in~\cite{KR08,KR09,KR10,KR10corr}.

{Summary:} The homotopic gluings for Platonic polyhedra from~\cite{EV04} are listed in Table~\ref{tab:homo}. We convert them into actions of 
deck groups $H$ on the covering three-sphere. 
They generate tilings by $|H|$ tiles of the covering three-sphere. 

\section{From Point Symmetry to Orbifolds}\label{sec:point}

The spherical Platonic polyhedra cover the three-sphere by deck transformations. 
By use of its point symmetry group $M$, we can 
decompose a~single Platonic polyhedron under $M$ into a~fundamental domain and its orbit. 
We must remove two obstacles towards a~topological interpretation:

(i) To keep the orientation under deck actions, we restrict the point group $M$ to proper rotations. For the Coxeter group action, its fundamental domain is a~simplex bounded by four Weyl reflection planes. When we restrict to rotations to preserve orientation,
the actions belong to the unimodular subgroup of $\Upgamma$ of determinant one, which we denote as $S\Upgamma$.
When passing to $S\Upgamma$, we must extend the fundamental simplex domain of $\Upgamma$ for $S\Upgamma$ to a~duplex, consisting of a~simplex and its image under a~reflection, see Figure~\ref{fig:duplices}.

(ii) A~topological manifold strictly excludes fixpoints under deck actions. Here, they 
appear on the rotation axes of the point groups. We must extend our topological notion from a~manifold 
to an~orbifold~\cite{TH97,MO87}, which allows for fixpoints of finite order.
The notion of orbifolds is explained in more detail in~\cite{RA94} (pp.~652--714). 
In Figure~\ref{fig:duplices}, we give the construction of orbifolds by gluing pairs of Coxeter 
simplices into duplices for the four Platonic polyhedra. The fixpoints appear on the edges of the 
blue glue triangles for the two Coxeter simplices.
 
To cover the full three-sphere by orbifolds, we proceed in two steps.
First of all, any Platonic polyhedron under its rotational point symmetry group $M$ is paved by $|M|$ copies of spherical orbifolds. 
These spherical orbifolds we take as duplices glued from a~Coxeter simplex and its mirror image. 
Copies of the duplices shown in Figure~\ref{fig:duplices} under the point group $M$ tile the Platonic polyhedra, but have fixpoints on their edges. To cover the full three-sphere, we augment the rotational point symmetry group $M$ by the operations of the full
deck group $H$ of the Platonic manifold. These two groups share only the identity element, $M\cap H=e$.
The product $M \cdot H$ of point and deck operations $m \in M, h \in H$, acting on a~proto-orbifold, generates 
as the deck group the unimodular subgroup $M\cdot H = S\Upgamma$ of the 
Coxeter group. The images of the proto-duplex under all of these products cover the three-sphere.

{Summary:} The topological orbifolds for Platonic spherical polyhedra are duplices (Figure~\ref{fig:duplices}), glued from two Coxeter simplices.
Their deck groups are the unimodular subgroups $S\Upgamma \in \Upgamma$. Each orbifold deck group is factorized as 
$S\Upgamma= M \cdot H,\: M\cap H= e$, into the 
point symmetry group $M$ and the deck group $H$ for the Platonic polyhedron.

\section{Harmonic Analysis for Orbifolds}\label{sec:orbi}

\begin{figure}[H]
\begin{center}
\includegraphics[width=0.51\textwidth]{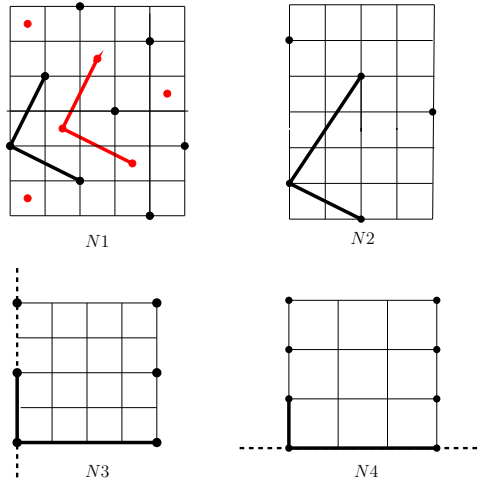}
\end{center}
\caption{\label{fig:subgrids}
Subgrids of integer/half-integer black/red points $(m_1 m_2)$ are spanned by heavy black/red vectors. The marked grid points select the bases $D^j_{m_1 m_2}$ for a~spherical orbifold $Ni$ from the square grid formed by thin lines. Coordinate transformations are given in Equation~(\ref{u14e}).
Tetrahedral case $N1$, coordinates $u(\tilde{x})$, integer black and half-integer red grid points.
Cubic case $N2$, coordinates $u(\tilde{x})$, with integer black grid points. 
Cubic case $N3$, coordinates $u(x)$, vertical broken mirror line, integer black grid points. 
Octahedral case $N4$, coordinates $u(x)$, horizontal broken mirror line, integer black grid points.}
\end{figure}

As the general basis for all of these topologies, we take the harmonic Wigner polynomials.
To adapt them to specific orbifolds, we shall present in the following subsections different routes. 
By graphical means, we shall illuminate on $2D$ $m$-subgrids; see Figure~\ref{fig:subgrids}, denumerable sets of Wigner polynomials for orbifolds, selected 
from the full harmonic set on the three-sphere. 
More details of the harmonic analysis on the Platonic orbifolds are given in Section~\ref{sec:harmonic}. 
\vspace{5pt}

\subsection{Rotation to the Diagonal Form}\label{sec:diag}

The unitary structure of the three-sphere allows one to reduce any given deck action to a~pure phase transformation.
A general rotation operator $T_{(g_l,g_r)},\: (g_l,g_r) \in SO(4,R)$ can be brought to a~simple form by unitary diagonalizations
of the matrices $(g_l,g_r)$, \vspace{-2pt}
\begin{eqnarray}
 \label{u14da}
&& g_l=q \updelta_l q^{-1},\: g_r= k \updelta_r k^{-1}, \quad q, k \in SU(2,C),
\\ \nonumber 
&&\updelta_l=\left[ 
\begin{array}{ll}
\exp(i\upalpha)& 0\\
0 & \exp(-i\upalpha)
\end{array}\right], 
\updelta_r=\left[ 
\begin{array}{ll}
\exp(i\upgamma)& 0\\
0 & \exp(-i\upgamma)
\end{array}\right] \vspace{4pt}
\end{eqnarray}
To find the angular parameters $\upalpha,\upgamma$ of the diagonal form, one can avoid the explicit diagonalization Equation~(\ref{u14da}).
For any $u \in SU(2,C)$, use the trace relation: \vspace{-2pt}
\begin{equation}
 \label{u14f} 
{\rm trace}(g_l)= 2\cos(\upalpha),\: {\rm trace}(g_r)= 2\cos(\upgamma) \vspace{-4pt}
 \end{equation}
to determine $(\upalpha,\upgamma)$.

We now pass from $u(x)$ to a~new matrix $u(\tilde{x})$ by the $SO(4,R)$ rotation: \vspace{-2pt}
\begin{equation}
\label{u14e}
q, k:\: u(x) \rightarrow u(\tilde{x})= q^{-1} u(x) k \vspace{-2pt}
\end{equation}
The matrix $u(\tilde{x})$ represents a~new system of coordinates on $S^3$.
It follows from Equation~(\ref{u14da}) that to the action of $T_{(g_l,g_r)}$ on the coordinates $u(x)$, there corresponds the action
$T_{(\updelta_l,\updelta_r)}$ on the coordinates $u(\tilde{x})$. If we pass in a~Wigner polynomial from the coordinates $\tilde{x}$
to Euler angles, we find in this parametrization: 
\begin{equation}
\label{u14q}
(T_{(\updelta_l,\updelta_r)} D^j_{m_1,m_2})(u(\tilde{x})) 
= D^j_{m_1,m_2}(u(\tilde{x}))\: \exp(i(-m_1\upalpha+m_2\upgamma))
\end{equation}
that is, a~pure phase transformation.
 The condition of invariance of a~Wigner polynomial $D^j_{m_1 m_2}(u')$ under the diagonal action of the operator $T_{(\updelta_l,\updelta_r)}$ 
now takes the form of a~phase condition: 
\begin{equation}
\label{xx1}
-m_1\upalpha+m_2\upgamma =\: \upnu\: 2\pi,\: \upnu= 0, \pm1, \pm2,... 
\end{equation}

{Summary:} Any deck action on~Wigner polynomials by a~rotation can be converted into a~pure phase transformation Equation~(\ref{u14q}). Any deck group with a~single-generator can be treated in this way. 

\subsection{Selection on m%italics or not? please check the convention throughout for math terms
-Subgrids}\label{sec:subgrids}

The harmonic analysis must be restricted to the modes admissible for a~specific spherical topology.
A denumerable set of Wigner polynomials $D^j(u)$ with the index pair $(m_1,m_2)$ 
and degree $2j, j \geq (|m_1|,|m_2|)$ is achieved by choosing from the harmonic Wigner basis 
a subbasis invariant under 
the relevant point and deck group. 
Since all deck actions are of finite order, the angles $\upalpha, \upgamma$ in Equation~(\ref{u14q}) must be integral fractions of $2\pi$ of the forms 
$\upalpha=p_1 \frac{2\pi}{q}, \upgamma=p_2\frac{2\pi}{q}$.
For groups with a~single generator, rewrite Equation~(\ref{xx1}) in the form:
\begin{eqnarray}
 \label{xx2} 
&&-m_1\upalpha+m_2\upgamma = \upnu\: 2\pi,\: \upalpha= p_1\frac{2\pi}{q}, \upgamma=p_2 \frac{2\pi}{q},
\\ \nonumber 
&& -m_1p_1+ m_2p_2= \upnu\: q,\: \upnu=0, \pm 1,\pm 2,\ldots \vspace{-2pt}
\end{eqnarray}
The solutions $(m_1,m_2)=(p_2,p_1)$ of Equation~(\ref{xx2}) for $\upnu=0$ form integer multiples of a~grid vector $a_1=(p_2,p_1)$. Solutions of Equation~(\ref{xx2}) for $\upnu \neq 0$ 
are on lines parallel to $a_1$. We choose a~second vector $a_2=(q_1,q_2)$ to a~grid point outside, but next to the line spanned by 
$a_1$. 
Then, the vectors $(a_1,a_2)$ form the basis of an~$m$-subgrid, whose points yield all solutions of Equations~(\ref{xx1}) and (\ref{xx2}).
This case applies to the orbifolds $N1$ and $N2$ with cyclic deck groups $C_5$, $C_8$.
In Figure~\ref{fig:subgrids}, we mark the subgrids for the orbifolds $N1$, $N2$, $N3$, $N4$.

Note that the topological rules Equation~(\ref{xx2}), given in terms of subgrids, select enumerable sets of harmonic Widner basis polynomials of unrestricted degree $2j$. 

\subsection{Mirror Extension of $m$-Grids}\label{sec:mirror}

The subgrid method can be extended to orbifolds when at most two 
Wigner polynomials with index pairs $[(m_1,m_2), (-m_1,m_2)]$ or $[(m_1, m_2), (m_1, -m_2)]$ 
linearly combine into an~invariant basis polynomial. We call these two cases vertical and horizontal mirror pairs. It then suffices to mark one~partner of this pair and to give the algebraic phase for its partner. These cases apply for the orbifolds $N2$ and $N3$, illustrated in Figure~\ref{fig:subgrids}.

\subsection{Projection of Harmonic Bases}\label{sec:projharmonic}

For more general deck groups, we establish the identity representation by application to a~Wigner polynomial 
$D^j$ of degree $2j$ the projection operator $P^0$ Equation~(\ref{u57}) to the identity representation of the deck group $H$, \vspace{8pt}
\begin{equation}
 \label{u57}
 P^0= \frac{1}{|H|} \sum_{g \in H} T_g\:\: \vspace{2pt}
\end{equation}
By Equation~(\ref{u14}), this application produces for any degree $2j$ a~linear combination of Wigner polynomials.
If this can be normalized, it contributes to the harmonic analysis. For Platonic spherical manifolds,
these projections were carried out in~\cite{KR10}. In many cases, it is possible to display the results 
of the projection Equation~(\ref{u57}) on an~$m$-subgrid: the non-vanishing Wigner polynomials selected by
projection are located on a~subgrid, dependent on the manifold, and spanned by two specific grid vectors $(a_1,a_2)$. We display subgrids in Figure~\ref{fig:subgrids}. 
The advantage of the subgrid method is that different topologies can be directly compared in terms of their subgrids. 

{Summary:} The Wigner polynomials invariant under a~single-generator deck group are located on a~subgrid spanned by two grid vectors $(a_1,a_2)$. Any subgrid point $(m_1,m_2)$ stands for an~enumerable set of invariant Wigner polynomials of degree $j=j_0+ \upnu,\: j_0=
max(|m_1|,|m_2|)$.
The spherical manifolds $N1, N2$ with cyclic deck groups $C_5, C_8$ belong to this set. 

\subsection{Algebraic Bases from Point Symmetry}

The basis of the harmonic analysis for a~specific topology can be recursively computed, as we shall show. We have seen that the deck group of the orbifold factorizes into the symmetry group $M$ of the Platonic manifold and its deck group $H$. The basis functions for the harmonic analysis must be invariant under the deck group $M \cdot H$, which is possible only if they are invariant under both $M$ and $H$. The two~projectors for the groups $M$ and $H$ commute~\cite{KR10c}. 

We start for the cubic orbifold $N3$ the first loop of a~recursive construction of basis functions of increasing degree $2j$:

(i) The action of symmetry groups $M=O$ of the Platonic manifolds is known from the Euclidean setting in $E^3$.
The point group $O$ is a~subgroup of the rotation group $SO(3,R)$; therefore, any $M$-invariants must appear for fixed $l$ as~linear combinations of 
spherical harmonics $\sum_m a_{lm} Y^l_m$. Standard spherical harmonics $Y_m^l$ with lowest label $l=l_0$ and invariant under $O$ can be found in~\cite{LA74} (pp.~436--438) for the cubic group and in~\cite{KG64} (pp.~105--114). For several point groups $M$, the onset $l_0$ of spherical harmonics invariant under $M$ is listed in~\cite{KR10a}, Table 3.
The linear combinations of lowest $l=l_0$ can be expressed in the spherical basis Equation~(\ref{u14b}) of $SO(4,R)$.
For the cubic point group $O$, the lowest linear combination of spherical harmonics given in Table~\ref{tab:cubicl} has $l_0=4$.

(ii) To also project on 
an invariant under $H$, which is $H=Q$ for the orbifold $N3$, we transform with Equation~(\ref{u14c}) from the spherical basis back to the Wigner basis and then apply the projector Equation~(\ref{u57})
with the matrix element:
\begin{equation}
\label{f11}
\langle j m_1'm_2'|\frac{1}{|H|}\sum_{(g_l,g_r)\in H} T_{(g_l,g_r)}|jm_1m_2\rangle 
\end{equation}
Once the result of (ii) is transformed back into the spherical basis with Equation~(\ref{u14b}), we recover linear combinations 
with new values of multipole order $l > l_0$, which, if non-vanishing, can be normalized. 
These new states must still be invariant under the point group $M$. 
It follows that they must contain $M$-invariant linear combination
of spherical harmonics with $l>l_0$. For the cubic point group $O$, we demonstrate this result in Table~\ref{tab:cubicl} with 
new cubic invariants for $l=6$ and $l=8$. 

With them, we enter the next loop with Steps (i) and (ii) in search of 
new invariant basis functions for the deck group $Q \times_s O$. This recursive method uses only well-known Wigner coefficients of $SU(2,C)$, but avoids any new projection to invariants under $M$.
The new basis functions in Table~\ref{tab:cubicinv} contain linear combinations of different multipole orders
$l=4, 6, 8$. 
This implies that topology can enforce linear correlations between amplitudes of different multipole order,
which has important implications for the analysis of the cosmic microwave background.

\begin{table}[H]
\centering
$
 \begin{array}{l|l}
l&Y^{\Upgamma_1,l}=\sum_m a_{lm}Y^l_m(\uptheta, \upphi)\\ \hline
0&Y^0_0\\
4&\sqrt{\frac{7}{12}}Y^4_0+\sqrt{\frac{5}{24}}(Y^4_4+Y^4_{-4})\\
6&\sqrt{\frac{1}{72}}Y^6_0-\sqrt{\frac{7}{144}}(Y^6_4+Y^6_{-4})\\
8&\frac{1}{64}\sqrt{33} Y^8_0+\frac{1}{12}\sqrt{\frac{21}{2}}(Y^8_4+Y^8_{-4})+
\frac{1}{24}\sqrt{\frac{195}{2}}(Y^8_8+Y^8_{-8})\\
\end{array}
$
\caption{\label{tab:cubicl} 
The lowest cubic invariant spherical harmonics $Y^{\Upgamma_1,l}$, expressed by spherical harmonics $Y^l_m$.}
\end{table}
\vspace{-12pt}
\begin{table}[H]
\centering
$
 \begin{array}{l|l|l}
2j&l&\uppsi^{0,\Upgamma_1,2j}=\sum_{l} b_{l} R_{2j+1\; l}(\upchi)Y^{\Upgamma_1,l}(\uptheta, \upphi)\\ \hline
0&0& R_{1 0} Y^{\Upgamma_1, 0}\\
4&0,4&\sqrt{\frac{2}{5}}R_{5 0}Y^{\Upgamma_1, 0}+\sqrt{\frac{3}{5}}R_{54}Y^{\Upgamma_1, 4}\\
6&0,4, 6&\sqrt{\frac{1}{7}}R_{7 0}Y^{\Upgamma_1, 0}-
\sqrt{\frac{6}{11}}R_{7 4}Y^{\Upgamma_1, 4}-\sqrt{\frac{24}{77}}R_{76}Y^{\Upgamma_1, 6}\\
8&0,4, 6, 8&\frac{4}{3}\sqrt{\frac{1}{110}}R_{9 0}Y^{\Upgamma_1, 0}
-\frac{12}{11}\sqrt{\frac{3}{65}}R_{9 4}Y^{\Upgamma_1, 4}\\
&&+\frac{8\cdot 19}{165}R_{96}Y^{\Upgamma_1, 6}
+\frac{4}{5}\sqrt{\frac{1}{33\cdot 13}}R_{98}Y^{\Upgamma_1, 8}\\
 \end{array}
$
\caption{\label{tab:cubicinv} The lowest $(Q \times_s O)$-invariant polynomials $\uppsi^{0,\Upgamma_1,2j}$
of degree $2j$ on the three-sphere for the orbifold $N3$, expressed by the cubic invariant spherical harmonics 
from Table \ref{tab:cubicl}.
$(Q \times_s O)$-invariance enforces coherent superpositions of several cubic invariant spherical~harmonics.
}
\end{table}

In Table~\ref{tab:cubicinv}, we exemplify the first basis functions for the cubic orbifold $N3$.

{Summary:} The recursive method, by use of loops with Steps (i) and (ii), yields a~basis of the harmonic analysis 
of any Platonic orbifold. Invariance under a~topological deck group will imply in general the correlation between multipoles of different orders $(l,l')$.

\section{Synopsis of Harmonic Bases for Platonic Orbifolds}\label{sec:harmonic}

Depending on the orbifold, we apply one of the methods given in the previous section to derive the basis of each Platonic orbifold.

\subsection{The Tetrahedral Orbifold $N1$}

The deck group of $N1$ from~\cite{KR10} is the cyclic group $H=C_5$ of order five. Its single generator is the~rotation: \vspace{3pt}
\begin{equation}
 \label{f12}
 T_{(W_1W_2W_3W_4)}= T_{(g_l, g_r)}: (g_l, g_r)=(v_1v_2^{-1}v_3v_4^{-1}, v_1^{-1}v_2v_3^{-1}v_4) \vspace{3pt}
\end{equation}
where the vector $a_i$ of the Weyl reflection $W_i := W_{a_i}$ of the Coxeter group from 
Table \ref{tab:coxb} by Equation~(\ref{b3}) determines
the unitary matrix $v_i= v(a_i), i=1,2,3,4$ (Equation~(\ref{b2})), and Equation~(\ref{b3}) is used to convert products of two Weyl reflection operators into rotations.

Equation~(\ref{f12}) demonstrates that the entanglement of left and right actions in the group $C_5$ is enforced by the geometry of the tetrahedron.

We follow Section~\ref{sec:wigner} and replace the old coordinates $u(x)$ by the new ones $u(\tilde{x})= q^{-1}u(x) k$. 
By the trace relation Equation~(\ref{u14f}) for $(g_l,g_r)$, we find the diagonal forms: \vspace{4pt}
\begin{equation}
\label{f13}
\updelta_l= \left[ 
\begin{array}{ll}
 \exp(i\: 2\pi/5)& 0\\
 0& \exp(-i\:2\pi/5)\\
 \end{array}\right],
 \:\:
\updelta_r= \left[ 
\begin{array}{ll}
 \exp(i\:6\pi/5)& 0\\
 0& \exp(-i\:6\pi/5)\\
 \end{array}\right] \vspace{3pt}
 \end{equation}
To project a~basis state invariant under $C_5$, it suffices to make it invariant under the generator Equation~(\ref{f12}) of $C_5$, since then, it will be invariant under any power of this generator.
We choose a~fixed Wigner polynomial and find in the coordinates $u(\tilde{x})$ as action the phase transformation:
\begin{equation}
 \label{f14}
(T_{(\updelta_l, \updelta_r)} D^j_{m_1,m_2})(u(\tilde{x})=D^j_{m_1,m_2}(\tilde{u})\exp(i(-m_12\pi/5+m_2\:6\pi/5) 
\end{equation}
To have invariance under the generator of $C_5$, we require: \vspace{-5pt}
\begin{equation}
\label{f15}
-m_1+3m_2 \equiv \: 0\: {\rm mod}\: 5 \vspace{-6pt}
\end{equation}
This condition is fulfilled on the subgrid points marked in Figure~\ref{fig:subgrids}, $N1$ by black and by red circles. 
The subgrids are spanned by the black vectors $a_1=(2,-1), a_2= (1,2)$ from an~integer or by the same red vectors from a~half-integer 
 grid point. 
The subgrid points form a~selection from all grid points.
The subbasis consists of all Wigner polynomials associated with the subgrid points.

\subsection{The Cubic Manifold $N2$}

The deck group is $C_8$ generated by: \vspace{2pt}
\begin{equation}
(g_l,g_r):
g_l= \left[\begin{array}{ll}
-\overline{a} & 0\\
0 & -a\\
\end{array} \right],\:\: 
g_r= \left[\begin{array}{ll}
0 &\overline{a}\\
 -a& 0\\
\end{array} \right],\: a=\exp(i\: 2\pi/8)
\end{equation}
Again, the left and right actions are entangled.

The subbasis of the harmonic analysis is of the horizontal mirror type (see Section~\ref{sec:mirror}) given
by the linear combination of two Wigner polynomials: 
\begin{eqnarray}
\Upphi^j_{m_1,0}&=& D^j_{m_1,0}(u),\: j \in Z, 
\\ \nonumber 
\Upphi^j_{m_1,m_2}&=& \frac{1}{\sqrt{2}}[ D^j_{m_1,m_2}(u)+ i^{(m_1+m_2)} (-1)^{(j+m_2)} D^j_{m_1,-m_2}(u)], 
\\ \nonumber
&& j \in Z,\: \text{for} \: m_1 \: \text{even},\: 0 <m_2\leq j
\end{eqnarray}

\subsection{The Cubic Manifold $N3$}

The deck group is the quaternionic group $Q$, acting from the left, with elements:
\begin{equation}\label{qg}
g_l=-{\bf k} = \left[ \begin{array}{ll}
  0&-i\\
  -i& 0\\
  \end{array}\right] ,\: 
g_l=-{\bf j} = \left[ \begin{array}{ll}
  0&-1\\
  1& 0\\
  \end{array}\right],\:  
g_l=-{\bf i}= \left[ \begin{array}{ll}
  -i&0\\
  0& i\\
  \end{array}\right]
\end{equation}
The subbasis is of vertical mirror type, given by a~linear combination of two Wigner polynomials:
\begin{eqnarray}
\Upphi^j_{m_1,m_2}&=& \frac{1}{\sqrt{2}}[D^j_{m_1,m_2}(u)- D^j_{-m_1,m_2}(u)]\\ 
&& \text{for}\:j\:\text{odd}, \geq 3, \text{for}\:m_1\:\text{even}, 0<m_1\leq j, -j\leq m_2\leq j
\end{eqnarray}

\subsection{The Octahedral Manifolds $N4$, $N5$, $N6$}

The deck groups here are of the order $48$. We give the basis construction to
the orbifold $N4$ with deck group $C^l_3 \times Q^r$ acting from the left and from the right, respectively.
	For the subbasis in new coordinates $u(\tilde{x})$, we find a~horizontal mirror symmetry and, from~\cite{KR10a} (p.~26), the following polynomials and mirror~phases: \vspace{-3pt}
\begin{eqnarray}
 m_1=\uprho \equiv 0\: {\rm mod}\: 3, && \nonumber\\ 
 \text{for}\:j\:\text{odd}, \geq 3,\: 0<m_2\leq j: 
 &\Upphi^j_{\uprho,m_2}&= [D^j_{\uprho, m_2}(u(\tilde{x}))-D^j_{\uprho,-m_2}(u(\tilde{x}))], \nonumber \\
 \text{for}\:j \:\text{even}, m_2=0:
 &\Upphi^j_{\uprho,0}&= D^j_{\uprho,0}(u(\tilde{x})), \nonumber \\
 \text{for}\:j \:\text{even}, \geq 2,\: 0< m_2 \leq j: 
 &\Upphi_{\uprho, m_2}&= [D^j_{\uprho, m_2}(u(\tilde{x}))+ D^j_{\uprho,-m_2}(u(\tilde{x}))]\nonumber \\ 
\end{eqnarray}
The octahedral manifolds $N5$, $N6$ are analyzed in~\cite{KR10a}.

\subsection{The Dodecahedral Manifold $N9$}

An analysis of invariant polynomials on Poincar\'{e}'s dodecahedral manifold is given in~\cite{KR05}, based on the work of Klein~\cite{KL93}.

\section{Topology of Multiply-Connected Universes}\label{sec:cmb}

As an~application of the derived subbases, we consider the observable effects of multiply-connected universes.
The topology of the cosmos has its roots in Albert Einstein's work~\cite{LU14}. 
In 1917, Einstein presented his pioneering paper on general relativity, cosmology and gravitation~\cite{EI17} (pp.~160--164). 
He communicates three fundamental ideas:

(i) {Space-time manifold and metric}: Space and time are unified into a~single 4D space-time manifold, with three space coordinates 
$(x_1,x_2,x_3)$ and one time coordinate $x_0=ct$. Following Riemann, this four-manifold carries a~pseudo-Euclidean metric with the space-time squared
distance: 
\begin{eqnarray}
\label{E1}
ds^2&=&\sum_{\upmu,\upnu=0}^3 g_{\upmu \upnu}(x_0,x_1,x_2,x_3)dx_{\upmu}dx_{\upnu}
\\ \nonumber 
&=& (ct)^2-\sum_{i,j=1}^3 g_{ij} dx_i dx_j,\\
g_{\upmu\upnu}&=&g_{\upnu\upmu}
 \end{eqnarray}
In the second line, we adopt the usual splitting into time and space. Space-time by Equation~(\ref{E1}) is distinct from $E^4$ used in 
previous sections.

(ii) {Gravity and geodesics}: Einstein gives field equations for Newton's gravity. They link the second derivatives of the 
metric tensor $ g_{\upmu \upnu}(x_0,x_1,x_2,x_3)$ with respect to the coordinates $x_{\upmu}$ linearly to the energy-momentum tensor 
$T_{\upsigma \uprho}(x_0,x_1,x_2,x_3)$. Geodesics, the shortest lines allowed by the metric Equation~(\ref{E1}), are the lines 
followed by massive test particles. For velocities small compared with the velocity $c$ of light, Einstein's field equations reduce to Newton's laws for the gravitational potential in the presence of masses and the differential equations for geodesics to Newton's equations of motion. In general, the metric and the energy-momentum tensor with Einstein become observables of astrophysics.

(iii) {Topology}: The connectivity of the space-time manifold is the subject of cosmic topology. The classes of possible closed paths are the basic concept of homotopy. The view of space-time as a~manifold, with gravitation obtained from local differential equations, presents topology as an~observable. 

When time is split off as in Equation~(\ref{E1}), homotopy refers to the $3D$ space part of space-time. 
In cosmology, three possible curvatures of the universe are distinguished: hyperbolic space with negative, Euclidean space with zero and spherical space with positive average curvature. 
The present astrophysical data favor positive curvature and, hence, a~spherical spatial topology, as discussed here.
One important observable is the spatial fluctuations of the cosmic microwave background.

\begin{figure}[H]
\begin{center}
\includegraphics[width=0.5\textwidth]{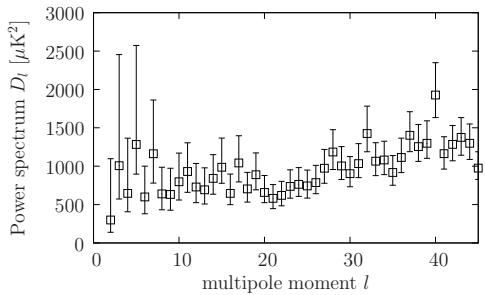} \vspace{-2pt}
\end{center}
\caption{\label{fig:cmb}
CMB temperature power-spectrum as a function of multipole order $l$; data from~\cite{PL13} (Figure~35).
Note the weak amplitudes of some of the lowest multipoles.}
\end{figure}

The cosmic microwave background (CMB) is an~observed uniform thermal black-body radiation of a~present temperature of 2.725 Kelvin with a~peak frequency of 160.2~GHz, discovered in 1964 by A.~Penzias and R. Wilson \cite{PW65}. According to big bang cosmology, its origin is dated back to the early epoch of photon decoupling, when neutral atoms were formed at very high density and temperature. 
Its amplitude today is observed by the Wilkinson probe and up to 2013 by the Planck satellite~\cite{PL13}. 
After a~non-trivial cleaning of the measured temperature for global astrophysical influences, such as the Wolfe--Sachs effect and the local foreground of the solar system and our galaxy,
the CMB spatial temperature fluctuations are expanded in multipoles, shown for the lowest values of $l$ in Figure~\ref{fig:cmb}. \vspace{8pt}

 \vspace{-4pt}
The connection of multipole-resolved CMB measurements to cosmic topology is provided by selection rules for the harmonic analysis.
The different topologies discussed here should imprint their selection rules on the multipole amplitudes of the CMB.
The cubic spherical manifold $N2$ discussed here has been used for detailed predictions of the CMB temperature fluctuation map~\cite{AKL11}.
In addition, the ratio of different multipole-moments encodes the specific topology; see Table~\ref{tab:cubicinv} for the cubic case.

Both topological signatures can be revealed with the methods given here, which yields regular tilings of the three-sphere.
According to~\cite{PL13}, Section~9.3 on p.~36, the observed low amplitudes and correlations of the first CMB multipoles (Figure~\ref{fig:cmb}) might indicate a~violation of statistical isotropy or might indicate the presence of topological selection rules.

Besides regular tilings, also random tilings of spherical three-manifolds have been analyzed in the literature~\cite{ES08}.

\section{Conclusions}\label{sec:conclusion}

We have constructed the orthonormal basis sets for four Platonic spherical topological orbifolds, 
required for performing a~harmonic analysis on specific topologies.
The mathematical framework used starts from the homotopy or fundamental groups and yields
isomorphic deck actions and deck groups on the three-sphere.
The point symmetry groups $M$ of the Platonic manifolds lead from manifolds to orbifolds.
The orthonormal Wigner polynomials provide a~basis for the harmonic analysis on $SU(2,C)$. 
By deriving the subgrid of each Platonic manifold on the space spanned by the indices of the Wigner polynomials,
we obtain selection rules for the harmonic analysis.
Moreover, the coefficients of the multipole expansion encode the underlying topology.

In contrast to other approaches, no numerical orthogonalization of basis sets is required, and
the connection between topology and geometry of space is transparent from the introduction of the 
homotopies.
Finally, we discussed the harmonic analysis of the CMB and present signatures of the underlying topology of
multiply-connected universes encoded in the CMB fluctuations.


\begin{thebibliography}{----}
\providecommand{\natexlab}[1]{#1}

\bibitem[1]{EI17}
Einstein, A.
\newblock {K}osmologische {B}etrachtungen zur allgemeinen
 {R}elativit{\"a}tstheorie.
\newblock In {\em Sitzungsberichte der K{\"o}niglich Preu{\ss}ischen Akademie der
 Wissenschaften (Berlin)}; {1917}; pp. 142--152.
% Publisher is "K{\"o}niglich Preu{\ss}ischen Akademie der Wissenschaften"
% full text online at: http://echo.mpiwg-berlin.mpg.de/MPIWG:H428RSAN

\bibitem[2]{KR05}
Kramer, P.
\newblock An~invariant operator due to {F} {K}lein quantizes {H}
 {P}oincar{\'e}'s dodecahedral 3-manifold.
\newblock {\em J. Phys. A Math. Gen.} {\bf 2005}, {\em
 38}, doi:10.1088/0305-4470/38/16/004.

\bibitem[3]{KL93}
Klein, F.
\newblock {\em Vorlesungen $\ddot{U}$ber das Ikosaeder}; Birkh{\"a}user: Basel, Switzerland, 1993.

\bibitem[4]{EV04}
Everitt, B.
\newblock 3-Manifolds from Platonic solids.
\newblock {\em Topol. Appl.} {\bf 2004}, {\em 138},~253--263.

\bibitem[5]{SE34}
Seifert, H.; Threlfall, W.
\newblock {\em Lehrbuch der Topologie}; Chelsea Reprint: New York, NY, USA, 1934.

\bibitem[6]{WO84}
Wolf, J.A.
\newblock {\em Spaces of Constant Curvature}, 5th ed.; Publish or Perish, Wilmington, DE, USA, 1984.

\bibitem[7]{TH97}
Thurston, W.P.
\newblock {\em Three-Dimensional Geometry and Topology}; Princeton University
 Press: Princeton, NJ, USA, 1997.

\bibitem[8]{RA94}
Ratcliffe, J.G.
\newblock {\em Foundations of Hyperbolic Manifolds}; Springer: Berlin, Germany, 1994.

\bibitem[9]{MO87}
Montesinos, J.M.
\newblock {\em Classical Tesselations and Three-Manifolds}; Springer: Berlin, Germany, 1987.

\bibitem[10]{KR10a}
Kramer, P.
\newblock Platonic topology and CMB fluctuations: Homotopy, anisotropy and
 multipole selection rules.
\newblock {\em Class. Quant. Grav.} {\bf 2010}, {\em 27}, doi:10.1088/0264-9381/27/9/095013.

\bibitem[11]{ED57}
Edmonds, A.R.
\newblock {\em Angular Momentum in Quantum Mechanics}; Princeton University
 Press: Princeton, NJ, USA, 1957.

\bibitem[12]{WI59}
Wigner, E.P.
\newblock {\em Group Theory and Its Application to the Quantum Mechanics of
 Atomic Spectra}; Wiley: New York, NY, USA, 1959.

\bibitem[13]{BA68}
Bargmann, V. Group representations on Hilbert spaces of analytic
 functions. 
\newblock In {\em Analytic Methods in Mathematical Physics}; Gordon and Breach:
 New York, NY, USA, 1968.

\bibitem[14]{LE02}
Levin, J.
\newblock Topology and the cosmic microwave background.
\newblock {\em Phys. Rep.} {\bf 2002}, {\em 365},~251--333.

\bibitem[15]{AU08}
Aurich, R.; Janzer, H.S.; Lustig, S.; Steiner, F.
\newblock Do we live in a~`small universe'?
\newblock {\em Class. Quant. Grav.} {\bf 2008}, {\em 25}, doi:10.1088/0264-9381/25/12/125006.

\bibitem[16]{MA66}
Magnus, W.; Karrass, A.; Solitar, D.
\newblock {\em Combinatorial Group Theory}; Dover: New York, NY, USA,~1966.

\bibitem[17]{HU90}
Humphreys, J.E.
\newblock {\em Reflection Groups and Coxeter Groups}; Cambridge University
 Press: Cambridge, UK, 1990.

\bibitem[18]{SO58}
Sommerville, D.M.Y.
\newblock {\em An~Introduction to the Geometry of N Dimensions}; Dover: New
 York, NY, USA, 1958.

\bibitem[19]{CM65}
Coxeter, H.S.M.; Moser, W.O.J.
\newblock {\em Generators and Relations for Discrete Groups}; Springer: New
 York, NY, USA, 1965.

\bibitem[20]{KR08}
Kramer, P.
\newblock Platonic polyhedra tune the 3-sphere: Harmonic analysis on simplices.
\newblock {\em Phys. Scr.} {\bf 2009},~{\em 79}, doi:10.1088/0031-8949/79/04/045008.

\bibitem[21]{KR09}
Kramer, P.
\newblock Platonic polyhedra tune the three-sphere: II. Harmonic analysis on
 cubic spherical three-manifolds.
\newblock {\em Phys. Scr.} {\bf 2009}, {\em 80}, doi:10.1088/0031-8949/80/02/025902.


\bibitem[22]{KR10}
Kramer, P.
\newblock Platonic polyhedra tune the three-sphere: III. Harmonic analysis on
 octahedral spherical three-manifolds.
\newblock {\em Phys. Scr.} {\bf 2010}, {\em 81}, doi:10.1088/0031-8949/81/02/025005.

\bibitem[23]{KR10corr}
Kramer, P.
\newblock Corrigendum: Platonic polyhedra tune the 3-sphere: Harmonic analysis
 on simplices.
\newblock {\em Phys. Scr.} {\bf 2010}, {\em 81}, doi:10.1088/0031-8949/81/1/019801.

\bibitem[24]{KR10c}
Kramer, P.
\newblock Multipole analysis in cosmic topology.
\newblock In \emph{Symmetries in Nature: Symposium in Memoriam Marcos Moshinsky}; American Institute of Physics (AIP)
 Publishing: College Park, MD, USA, 2010; Volume 1323, pp. 164--177.

\bibitem[25]{LA74}
Lax, M.
\newblock {\em Symmetry Principles in Solid State and Molecular Physics};
 Wiley: New York, NY, USA,~1974.

\bibitem[26]{KG64}
Knox, R.S.; Gold, A.
\newblock {\em Symmetry in the Solid State}; W.A. Benjamin: New York, NY, USA, 1964.

\bibitem[27]{LU14}
Luminet, J.P.
\newblock Cosmic topology: Twenty years after.
\newblock {\em Gravit. Cosmol.} {\bf 2014}, {\em 20},~18--20.

\bibitem[28]{PW65}
Penzias, A.A.; Wilson R.
\newblock Measurement of excess antenna temperature at 4080 Mc/s.
\newblock {\em Astrophysical Journal} {\bf 1965}, {\em
 142},~419--421.
 
\bibitem[29]{PL13}
Planck~Collaboration.
\newblock Planck 2013 results. XV. CMB power spectra and likelihood.
\newblock {\em Astron. Astrophys.} {\bf 2013}, {\em 571},~A15.

\bibitem[30]{AKL11}
Aurich, R.; Kramer, P.; Lustig, S.
\newblock Cosmic microwave background radiation in an~inhomogeneous spherical
 space.
\newblock {\em Phys. Scr.} {\bf 2011}, {\em 84}, doi:10.1088/0031-8949/84/05/055901.

\bibitem[31]{ES08}
Escudero, J.G.
\newblock Random tilings of spherical 3-manifolds.
\newblock {\em J. Geom. Phys.} {\bf 2008}, {\em
 58},~1451--1464.

\end{thebibliography}
\end{document}